\documentclass[%
 reprint,
 amsmath,amssymb,
 aps,
floatfix,
]{revtex4-2}
\usepackage{graphicx}
\usepackage{dcolumn}
\usepackage{bm}
\usepackage{physics}
\usepackage{caption}
\usepackage{subcaption}

\usepackage{amsmath}
\usepackage{tensor}
\usepackage{dsfont}

\newcommand{\levicivita}{}
\def\levicivita#1#{\tensor#1{\epsilon}}
\usepackage[hidelinks,colorlinks=true,linkcolor=blue,citecolor=blue, filecolor=blue, urlcolor=blue]{hyperref}

\begin{document}

\preprint{APS/123-QED}

\title{Transforming spin chains with a continuous driving field}

\author{Hudaiba Soomro, Adam Zaman Chaudhry}
 \altaffiliation[]{}
 \email{adam.zaman@lums.edu.pk}
\affiliation{%
 School of Science and Engineering, Lahore University of Management Sciences (LUMS),
Opposite Sector U, D.H.A, Lahore 54792, Pakistan
}%
\date{\today}

\begin{abstract}

A continuous, sinusoidal control field is used to suitably transform quantum spin chains. In particular, we are able to transform the quantum Ising chain to the quantum XY model, and the XY model to the XYZ spin chain. Our applied control field can also mitigate the effect of noise on the spin chain. We show how these spin chain transformations can be useful for quantum state transfer as well as entanglement generation.

\end{abstract}

\maketitle
\section{\label{sec:level1}Introduction\protect}

Quantum spin chains have served as prototypical models in explaining a host of physical phenomena. For instance, spin chain models have been used to study many body localization \cite{Ref2, Ref3}, Anderson localization, quantum chaos \cite{Ref4}, phase transitions \cite{Ref5, Ref6}, and quantum correlations \cite{Ref7}. Their theoretical prowess has paved the way for their experimental realization in the form of optical lattices \cite{Ref8}, ion traps such as Paul traps \cite{Ref9, Ref10}, solid state setups such as nitrogen vacancy centers connected by a chain of implanted nitrogen impurities \cite{Ref11, Ref12}, as well as photonic systems \cite{Ref13}. Moreover, numerical simulations also assist research directions, especially for complex spin chains which unlike low dimensional spin chains (1D and 2D) are not exactly solvable \cite{Ref1}. For instance, Ref.~\cite{Ref14} uses a transformation scheme to simulate the transport of polarization beyond exactly solvable methods. Similarly, a numerical pulse optimization technique and a unitary operator decomposition algorithm to simulate unitary operators in spin chains. Furthermore, quantum simulators allow the possibility of simulating spin chains using quantum circuits to study phenomena such as phase transitions in the thermodynamic limit \cite{Ref16}.

In the field of quantum information, quantum spin chains have yielded extremely useful results, from accurate state transfer to generation and transfer of entangled states \cite{Ref17, Ref18, Ref19, Ref20}. Experimental possibilities of attaining perfect state transfer have included implementations such as hybrid integrated quantum photonic circuits and superconducting qubits \cite{Ref27, Ref28}. Theoretically, there are various methods to achieve perfect state transfer in spin chains \cite{Ref21, Ref22, Ref23, Ref24, Ref25}. One particularly useful method works by modulating neighboring coupling strengths to achieve perfect state transfer in the XY spin chain \cite{Ref26}. Similarly, entanglement improvement in quantum spin system is another crucial site of work. While there are methods of improving long-distance entanglement such as the use of quantum mediators - for example, sets of antiferromagnetically coupled spin dimers \cite{Ref29} - we focus on the use of control field configurations \cite{Ref30, Ref31, Ref32}. The use of control fields has a dual advantage. First, they can be used to perform useful Hamiltonian transformations. For instance, in some cases, the use of oscillatory and static control fields can effectively introduce additional spin-spin interaction terms which can in turn improve entanglement generation \cite{Ref33, Ref34, Ref35, Ref36}. Similarly, a staggered field configuration has been shown to produce perfect entanglement between any two spin sites of an XY spin chain \cite{Ref37}. Second, control fields can be used to dynamically decouple the system from the environment so that the effect of noise on the spin chain is mitigated \cite{Ref38, Ref39, Ref40, Ref41, Ref42}, thereby helping to protect the fragile quantum states from decoherence. More so, the method of dynamical decoupling can lead to additional benefits such as scaling of the squeezing performance in spin chains \cite{Ref43}. 

Keeping in mind the usefulness of suitable control fields, we propose the use of a continuous driving field that can effectively transform spin-spin interactions. In particular, our work is inspired by the use of a continuous driving field to transform a one-axis spin squeezing Hamiltonian can be transformed to a two-axis spin squeezing Hamiltonian \cite{Ref44}. Using a similar scheme, we show how we can effectively transform the quantum Ising spin chain to the XY spin chain, and the XY spin chain to an engineered XYZ spin chain. These spin chain transformations introduce additional interactions in the spin chain Hamiltonian which can be controlled by tuning the control field parameters. Suitably engineering the spin-spin interactions allows us to yield perfect fidelity of state transfer for the quantum Ising spin chain system with a control field as this transformation allows us to realize the isotropic XY spin chain \cite{Ref28}. Similarly, by our control fields effectively transform the XY spin chain such that significantly higher entanglement is generated between the ends of the spin chain. Moreover, while we observe no significant entanglement in the Ising spin chain, we are able to show that by taking into account the next nearest-neighbor interactions and subjecting the spin chain to a transverse control field, considerably greater entanglement can again be generated between the ends of the spin chain. Apart from suitably engineering the spin-spin interactions, the highly oscillatory control fields are also useful in averaging out system-environment interaction terms which manifest as local noise terms and lead to decoherence of the fragile quantum states \cite{Ref41}.

\section{The Theoretical Model}

\subsection{Transforming the Ising spin chain}

The Hamiltonian for the quantum Ising spin chain can be writted as (we use dimensionless units throughout with $\hbar = 1$) 
\begin{equation}
    H = \sum_{i=1}^{N-1}J_{i}\sigma_x^{i} \sigma_x^{i+1},
\label{eq1}
\end{equation}
where $i$ labels the sites, and $J_i$ describe the nearest-neighbor coupling strengths. In the presence of a sinusoidal time-dependent field in a transverse direction, the Hamiltonian becomes 
\begin{equation}
    H(t) = \sum_{i=1}^{N-1}J_{i}\sigma_x^{i} \sigma_x^{i+1} + \sum_{i=1}^{N}h(t)\sigma_z^{i},
\label{eq2}
\end{equation}
where $h(t)= g\cos(\omega t)$, with $g$ being the strength of the driving field, and $\omega$ its frequency. Since the Hamiltonian at different times does not commute, finding the unitary time-evolution operator is far from straightforward. To make progress, it is best to transform to the frame of the applied field, that is, we define \cite{Ref44}
\begin{equation}
    H_{I}(t) = U_{c}^{\dagger}(t)VU_{c}(t),
\label{eq3}
\end{equation}
where $V$ is the time-independent part of the Hamiltonian ($\sum_{i=1}^{N-1}J_{i}\sigma_x^{i} \sigma_x^{i+1}$) and $U_{c}(t)$ is found as
\begin{equation}
\begin{split}
    U_{c}(t)=&\exp[-i\int_{0}^{t}\sum_{i=1}^{N} g\cos(\omega\tau)\sigma_z^{i}d\tau]\\
            =&\exp[-i\sum_{i=1}^{N}g \frac{\sin(\omega t)}{\omega}\sigma_z^{i}].
            \end{split}
\label{eq4}
\end{equation}
Defining $C(t)=\frac{\sin(\omega t)}{\omega}$ for simplicity of notation, we get
\begin{equation}
    H_{I}(t)=e^{igC(t)\sum_{i=1}^{N}\sigma_z^{i}}\left(\sum_{i=1}^{N-1}J_{i}\sigma_x^{i}\sigma_x^{i+1}\right)e^{-igC(t)\sum_{i=1}^{N}\sigma_z^{i}}.
\label{eq5}
\end{equation}
Introducing the raising and lowering operators in the usual fashion, that is, $\sigma_{\pm}^j= \sigma_{x}^j\pm i\sigma_{y}^j$, we note that 
\begin{equation}
\label{A1}
e^{igC(t)\sum_{j=1}^N\sigma_z^j}\sigma_{\pm}^{j}e^{-i gC(t)\sum_{j=1}^{N}\sigma_z^j} = \sigma_{\pm}^{j}e^{\pm2igC(t)}.
\end{equation}
The Hamiltonian in Eq.~\eqref{eq5} then transforms as
\begin{eqnarray}
H_{I}(t)&=\frac{1}{4}[\sum_{i=1}^{N-1}J_{i}(e^{4ig C(t)}\sigma_{+}^{i}\sigma_{+}^{i+1} +e^{-4ig C(t)} \sigma_{-}^{i}\sigma_{-}^{i+1})+\nonumber\\
&J_{i}(\sigma_{+}^{i}\sigma_{-}^{i+1} + \sigma_{-}^{i}\sigma_{+}^{i+1})]
\label{A2}
\end{eqnarray}
The Hamiltonian in the frame of the applied field has split into a time-dependent part and a time-independent part. We now make use of the Jacobi-Anger expansion \cite{stegun}, namely
\[e^{\frac{i4g\sin(\omega t)}{\omega}}=\sum_{n=-\infty}^{\infty} \mathcal{J}_n \left(\frac{\pm4g}{\omega}\right)e^{i n\omega t},\]
where $\mathcal{J}_n$ is the $n^{\text{th}}$ Bessel function of the first kind. Note that until now, our treatment has been exact. We now assume that the applied field has a high frequency such that $\omega \gg J_{i}$. In such a scenario, the higher order terms in the Jacobi-Anger expansion can be ignored, and we can consider only the $n = 0$ term. Defining $A = \mathcal{J}_0(\pm \frac{4g}{\omega})$, we achieve the time-independent effective Hamiltonian
\begin{equation}
    H_{I}= \sum_{i}^{N-1}\frac{J_i}{2}[(A+1)\sigma_{x}^{i}\sigma_{x}^{i+1}-(A-1)\sigma_{y}^{i}\sigma_{y}^{i+1}]. 
    \label{eq6}
\end{equation}
Since this effective Hamiltonian is time-independent, we can find the corresponding time-evolution operator $U_I(t) = e^{-iH_It}$. The full time-evolution operator is then given by $U(t) = U_c(t)U_I(t)$. However, the time-evolution operator $U_c(t)$ is periodic with period $2\pi/\omega$. As such, at times equal to integer multiples of $2\pi/\omega$, the time evolution generated by the full Hamiltonian is simply given by $U_I(t)$, provided of course that $\omega \gg J_i$. In particular, starting from the quantum Ising model, it is clear that we have effectively transformed to the XY spin chain via the applied control field. The anisotropy for this effectively spin chain can be changed by tweaking the control field's parameters. For instance, taking $A=0$ yields the fully isotropic XY spin chain - this entails requiring that $\frac{4g}{\omega}$ equals the roots of the zeroeth Bessel function. On the other extreme, if we choose $A=1$, we recover the initial Ising spin chain. It should be noted that there is a restriction on the range of values $A$ can take, namely $-0.4\leq A \leq 1$; this simply follows from the properties of the Bessel function. 

\label{Sec_2_a_1}

\subsection{Transforming the XY spin chain}

\label{Sec_2_a_2}
The transformation that we have discussed can analogously be extended to transforming the XY spin chain. We begin with the Hamiltonian 
\begin{eqnarray}
H(t)&= \sum_{i}^{N-1}[J_{x}^{i}\sigma_{x}^{i}\sigma_{x}^{i+1}+J_{y}^{i}\sigma_{y}^{i}\sigma_{y}^{i+1}]+\nonumber\\
&\sum_{i=1}^N[g\cos({\omega t})\sigma_z^i].
\label{eq7}
\end{eqnarray}
As before, we transform to the interaction picture where the Hamiltonian transforms as $H_I(t)= U_c^{\dagger}(t)VU_c(t)$ where $V$ is the time-independent part of the Hamiltonian in Eq.~(\ref{eq7}). As in Eq.~(\ref{eq4}), we find $U_c(t)$ is given by 
$U_c(t)= \exp[-i\sum_{i=1}^{N}gC(t)\sigma_z^{i}]$, with $C(t) = \frac{\sin(\omega t)}{\omega}$ as defined previously. We then get
\begin{eqnarray}
H_I&=\sum_{i}^{N-1}[\frac{1}{2}\{(J_{x}^{i}(A+1)-J_{y}^{i}(A-1))\sigma_{x}^{i}\sigma_{x}^{i+1}+\nonumber\\
&(J_{y}^{i}(A+1)-J_{x}^{i}(A-1))\sigma_{y}^{i}\sigma_{y}^{i+1}\}],
\label{eq9}
\end{eqnarray}
where, as before, $A = \mathcal{J}_0(\pm \frac{4g}{\omega})$. We see that if we choose $J_x^i = J_y^i$, we get back the fully isotropic XY spin chain. Our chosen control field is unable to yield additional interaction terms. To get these, let us consider the more interesting case where we apply the field along one of the two axes of the Hamiltonian. The term along the same axis does not undergo any transformation while the term along the transverse axis transforms yielding an interaction term along the third axis. As such, we now apply a control field to the XY spin chain along the $y$ direction, that is, we consider
\begin{eqnarray}
H(t)&= \sum_{i}^{N-1}[J_{x}^{i}\sigma_{x}^{i}\sigma_{x}^{i+1}+J_{y}^{i}\sigma_{y}^{i}\sigma_{y}^{i+1}]+\nonumber\\
&\sum_{i=1}^N[g\cos({\omega t})\sigma_y^i]
\label{eq10}
\end{eqnarray}
Carrying out an analogous procedure as done before, but now with $U_c(t) = \exp[-i\sum_{i=1}^{N}gC(t)\sigma_y^{i}]$, we get
\begin{eqnarray}
H_I&=\sum_{i}^{N-1}[\frac{1}{2}\{J_{x}^{i}(A+1)\sigma_{x}^{i}\sigma_{x}^{i+1}-J_{x}^{i}(A-1)\sigma_{z}^{i}\sigma_{z}^{i+1}\nonumber\\
&+2J_{y}^{i}\sigma_{y}^{i}\sigma_{y}^{i+1}\}].
\label{eq12}
\end{eqnarray}
Taking $A=0$, $J_i^x=2$, and $J_i^y=0$ yields the XXX spin chain given by 
\begin{eqnarray}
H_I&=\sum_{i}^{N-1}[\sigma_{x}^{i}\sigma_{x}^{i+1}+\sigma_{y}^{i}\sigma_{y}^{i+1}+\sigma_{z}^{i}\sigma_{z}^{i+1}].
\label{eq13}
\end{eqnarray}
Another useful case is obtained if we choose $J_x^i=J_y^i=J_i$ and $A=0$ in Eq.~(\ref{eq12}). We then obtain
\begin{eqnarray}
H_I&=\sum_{i}^{N-1}[\frac{J_i}{2}\{\sigma_{x}^{i}\sigma_{x}^{i+1}+\sigma_{z}^{i}\sigma_{z}^{i+1}+2\sigma_{y}^{i}\sigma_{y}^{i+1}\}],
\label{eq14}
\end{eqnarray}
which is essentially a rotated version of the usual XXZ spin chain.  

We can also recast our results for the XY spin chain in terms of an anisotropy parameter $\gamma$. Let us rewrite our spin chain Hamiltonian in the presence of the control field as 
\begin{eqnarray}
H(t)&=\sum_{i}^{N-1}J_i[(\gamma+1)\sigma_{x}^{i}\sigma_{x}^{i+1}+(1-\gamma)\sigma_{y}^{i}\sigma_{y}^{i+1}]\nonumber\\
&+\sum_{i=1}^N[g\cos({\omega t})\sigma_y^i]
\label{eq15}
\end{eqnarray}
We correspondingly have that 
\begin{eqnarray}
H_{I}&=\sum_{i}^{N-1}\frac{J_i}{2}[(\gamma+1)(A+1)\sigma_{x}^{i}\sigma_{x}^{i+1}-\nonumber\\
&(\gamma+1)(A-1)\sigma_{z}^{i}\sigma_{z}^{i+1}+2(1-\gamma)\sigma_{y}^{i}\sigma_{y}^{i+1}].
\label{eq16}
\end{eqnarray}
\subsection{Introducing Noise}
\label{Sec_2_c}

An experimentally realizable model of any quantum system must take into account noise introduced due to the environment. Assuming that each spin locally interacts with the environment, the system-environment interaction can be modelled as follows: 

\begin{equation}
    H_{SB}=\sum_{i=1}^N (B_x^i\sigma_x^i +B_y^i\sigma_y^i + B_z^i\sigma_z^i).
\label{eq17}
\end{equation}

In Eq.~(\ref{eq17}), $B_m^i$ represent the randomly fluctuating noise terms; these could even be environment operators. In order to eliminate these system-environment interaction terms, we use the method of dynamically decoupling the open quantum system by means of a control field \cite{Ref39}. Since our control field acts along one of the axes, it is able to cancel out noise terms along two of the three general degrees of freedom \cite{Ref40}. This is because $U_c(t)$ is a rotation operator and causes the noise terms that are perpendicular to it to rotate such that they average out. However, the noise term which is placed along the same axis as $U_c(t)$ remains undisturbed. Examining this in more detail, we note that in order for the periodic control field to decouple the noise terms, we need \cite{Ref38, Ref39, Ref40, Ref41, Ref42}
\begin{equation}
\int_0^{t_c} U_c^\dagger(t) H_{SB} U_c(t) dt=0.
\label{eq18}
\end{equation}
For the control field in the $z$ direction, we have $U_c(t)=\exp(-i\sum_{i=1}^{N}g \frac{\sin(\omega t)}{\omega}\sigma_z^{i})$. We then consider
\begin{eqnarray*}
\int_0^{\frac{2\pi}{\omega}} e^{i\sum_{i=1}^{N}g \frac{\sin(\omega t)}{\omega}\sigma_z^{i}}\sum_{i=1}^N B_x^i\sigma_x^ie^{-i\sum_{i=1}^{N}g \frac{\sin(\omega t)}{\omega}\sigma_z^{i}}\\
+e^{i\sum_{i=1}^{N}g \frac{\sin(\omega t)}{\omega}\sigma_z^{i}}\sum_{i=1}^{N}B_y^i\sigma_y^i e^{-i\sum_{i=1}^{N}g \frac{\sin(\omega t)}{\omega}\sigma_z^{i}}dt.
\label{B2}
\end{eqnarray*}
To solve this, we note that 
\begin{eqnarray*}
e^{iC(t)\sum_{i=1}^N \sigma_x^i} \sigma_y^j e^{-iC(t)\sum_{i=1}^N \sigma_x^i}= \\
\sigma_y^j\cos(2C(t)) -\sigma_x^j\sin(2C(t)).
\label{B3}
\end{eqnarray*}

It is straightforward to see that integrating over $\sin{(2C(t))}$ yields zero while integrating over $\cos{(2C(t))}$ yields the zeroth Bessel functions of the first kind being divided by $\omega^2$. As the zeroth Bessel functions are bounded by $1$, increasing $\omega$ leads to the integral approaching zero. Having analytically satisfied the decoupling condition, we will model, for the purposes of the ensuing numerical simulations, the noise as Ornstein-Uhlenbeck processes keeping the mean ($\mu=0$), standard deviation ($\sigma=0.5$), and the correlation time ($\tau=0.005$) \cite{Ref50}. 

\section{Numerical Results}
\label{result_section}

In this section, we present simulation results that help illustrate the usefulness of the spin chain transformations from the point of view of quantum information processing tasks. We present these results for the following effectively transformed time-independent spin chain systems: (i) the isotropic XY spin chain, (ii) the XXX spin chain, (iii) the rotated XXZ spin chain, and (iv) the XY spin chain with next nearest neighbor interactions. For the first case, we present the fidelity of state transfer as well entanglement generation while for the rest of the cases, we present the concurrence of entanglement generation between the ends of the spin chain, thereby showing that one can effectively modulate the interaction couplings to enhance entanglement generation.

\subsection{The isotropic XY spin chain}

\subsubsection{Quantum State Transfer}

We first consider the isotropic XY spin chain which we get as the effective spin chain after applying a transverse field to the Ising spin chain, 
\begin{equation}
    H_{I}(t)= \sum_{i}^{N-1}\frac{J_i}{2}[\sigma_{x}^{i}\sigma_{x}^{i+1}+\sigma_{y}^{i}\sigma_{y}^{i+1}]. 
    \label{eq19}
\end{equation}
This requires that we set $A=0$ in Eq.~(\ref{eq6}). From the definition of $A=\mathcal{J}_0(\frac{4g}{\omega})$, we observe that $A=0$ requires that ($\frac{4g}{\omega}$) equals the roots of the zeroeth Bessel function. We use the first of these zeroeth roots in our simulations ($g=\frac{\omega}{4(2.4048)}$) for the corresponding time-dependent spin chain system described by Eq.~(\ref{eq2}). Moreover, it is important to note that in deriving the result in Eq.~(\ref{eq19}), we assumed that $\omega \gg J_i$. As such, we use a rapidly oscillating field throughout the simulations. Having decided the control field's parameters, we initialize the spin chain sites in the state $\ket{\Psi(0)}=(\alpha\ket{0}_z + \beta\ket{1}_z) \otimes\ket{1}_z \ldots\otimes\ket{1}_z$. For quantum state transfer, after some time $\tau$, the state should evolve to $\ket{\Psi(\tau)}=\ket{1}_z\otimes\ket{1}_z\ldots \otimes (\alpha\ket{0}_z + \beta e^{i\phi}\ket{1}_z)$, where $\phi$ is known and can be corrected at the end of the state transfer \cite{Ref33}. We study the state transfer $\ket{\Psi_i}=\ket{0}_z\otimes\ket{1}_z\ldots\otimes\ket{1}_z \rightarrow \ket{\Psi_f}=\ket{1}_z\otimes\ket{1}_z\ldots\otimes\ket{0}_z$; by showing that we can effectively perform this state transfer, we can show that we can perform state transfer for an arbitrary superposition state. To quantitatively measure the state transfer, we find the fidelity, $F(t)= \bra{\psi_f}\rho(t)\ket{\psi_f}$, where $\rho(t)$ is the spin density matrix as a function of time. 

\begin{figure}[h!]
    \centering
    \includegraphics[width=0.35\textwidth]{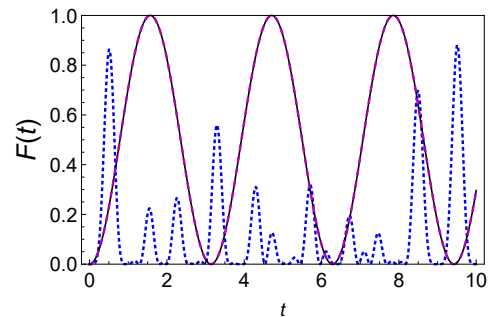}
    \caption{The smooth black curve shows the fidelity of state transfer for the Ising spin chain in the presence of a transverse field for $N = 7$. The dashed magenta curve shows the fidelity in the time-effective XY spin chain while the dotted blue curve is for fidelity in the Ising spin chain without any fields. Here, we have set the neighboring coupling strengths as $J_i=\sqrt{i(N-i)}$.}
    \label{fig1}
\end{figure}

For an isotropic XY spin chain, it has been shown that perfect fidelity of state transfer can be achieved by engineering the coupling interaction $J_i$ as $\sqrt{i(N-i)}$ \cite{Ref26}. As such, we select the nearest neighbor coupling terms as $J_i = \sqrt{i(N-i)}$ and simulate the fidelity plots corresponding to Eq.~(\ref{eq19}) and its time-dependent equivalent. These are described the dashed magenta curve and the smooth black curve respectively for $N=7$ in Fig.~(\ref{fig1}) which are shown to overlap. In Fig.~(\ref{fig1}), the blue dotted curve is for the fidelity of the Ising spin chain in the absence of any fields showing the relative advantage of using the Ising spin chain with the control field. Moreover, we have investigated that the fidelity for higher $N$ values further falls for the Ising spin chain alone while it remains practically perfect in the presence of the control field.

Having verified the behaviors of the time-dependent as well as the corresponding time-independent spin chain, we now introduce noise into the system to see how closely the behavior of our noisy time-dependent spin chain system matches with the effective noiseless case. As discussed before, we consider the noise terms to be given by
\[H_{SB}=\sum_{i=1}^N (B_x^i\sigma_x^i +B_y^i\sigma_y^i),\]
since these noise terms can be removed with our control field. We regard the noise terms $B_{m}^i$ as independent random variables that are found by solving the Ornstein-Uhlenbeck equation
\[dB_{m}= -\frac{(B_{m}-\mu)}{\tau}dt+\sigma\sqrt{\frac{2}{\tau}}dW,\]
where $m=x,y$, while $\mu$ is the mean, $\sigma$ is the standard deviation, $\tau$ is the correlation time, and $W$ is the standard Weiner process \cite{Ref50}. Here, $\mu$ and $\sigma$ have the same dimensions as the $B_m$. For all the simulations dealing with noise in this paper, we have kept $\sigma = 0.5$, $\mu = 0$, and $\tau = 0.005$ while sampling the noise over $20$ trials.

\begin{figure}[h!]
    \centering
    \includegraphics[width=0.35\textwidth]{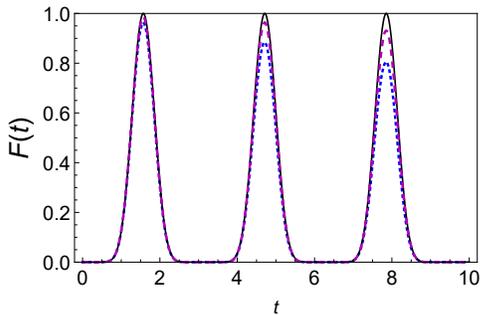}
    \caption{The smooth black curve here shows the fidelity of state transfer for an Ising spin chain in the presence of a transverse field for $N=7$. The dashed magenta curve is for the Ising spin chain with both the field and the noise while the dotted blue curve is for the XY spin chain in the presence of noise. Here, we have set the neighboring coupling strengths as $J_i=\sqrt{i(N-i)}$.}
    \label{fig2}
\end{figure}

As explained before, we expect that as a result of the decoupling condition being met, the noise simulations should match closely with the ones without noise. As such, the fidelity simulation for the Ising spin system with the control field in the $z$ direction for $N=7$ is shown in Fig.~(\ref{fig2}) by the solid black curve while the dashed magenta curve reveals the effect of noise. Furthermore, we also plot the XY spin chain (no control fields) in the presence of the same noisy environment and find that the fidelity for the XY spin chain and noise dips more rapidly as compared to the Ising spin chain with the transverse field in the presence of noise. 

\begin{figure}[h!]
    \centering
    \includegraphics[width=0.35\textwidth]{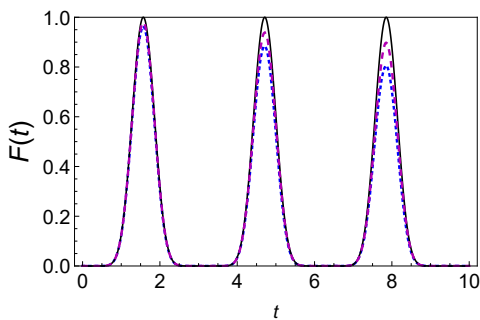}
    \caption{The smooth black curve shows the fidelity of state transfer of the XY spin chain in the presence of a control field along the $z$ direction for $N=7$. The dashed magenta curve accounts for the XY spin chain with the field in the $z$ direction in the presence of noise while the dotted blue curve is for the initial XY spin chain with just noise. Here, as usual, we have set $J_i=\sqrt{i(N-i)}$}
    \label{fig3}
\end{figure}

As a digression, let us consider the case of the isotropic XY spin chain upon which is incident a control field in the $z$ direction. Earlier, in Eq.~(\ref{eq9}), we had shown that this does not transform the spin chain. Let us now demonstrate this by simulating the fidelity of state transfer in such a system. Setting $J_x^i=J_y^i=J_i=\sqrt{i(N-i)}$, we plot the fidelity of state transfer for the XY spin chain in the presence of the control field (the black curve) in Fig.~(\ref{fig3}); the dashed magenta curve shows the effect of noise on the same system. As is the case for the XY spin chain demonstrated in Fig.~(\ref{fig2}), the fidelity remains practically perfect, hence verifying that the transformation has not affected the spin chain. Moreover, the control field cancels out the effect of noise as we observe the magenta curve to closely follow the the smooth black curve. Moreover, the dotted blue curve in Fig.~(\ref{fig3}) is for just the XY spin chain in the presence of noise (that is, no control fields) which we observe dips faster than the magenta curve.

\subsubsection{Quantifying Entanglement Generation}
In this section, we turn to quantify the generation of entanglement between two spin sites in a spin chain by finding the concurrence between the two spins. We do so for an initially bit flipped state $\ket{\psi_i} = \ket{0}_z\otimes\ket{1}_z\ldots\otimes\ket{1}_z$. First, in the time evolved spin density matrix, we take the trace over all the spins other than the two being considered. After finding this density matrix, $\rho'(t)$, we find the concurrence by first finding
\[R= \sqrt{\sqrt{\rho'}\widetilde{\rho'}\sqrt{\rho'}},\]
where
\[\widetilde{\rho'}=(\sigma_y\otimes\sigma_y)\widetilde{\rho '}^{*}(\sigma_y\otimes\sigma_y).\]
The concurrence, $C(t)$ is then given by
\[C(\rho)=\textrm{max}(0, \lambda_1 -\lambda_2 - \lambda_3 - \lambda_4),\]
where each of the $\lambda_i$ above are eigenvalues of $R$ in descending order \cite{Ref33}. 

\begin{figure}[h!]
    \centering
    \includegraphics[width=0.35\textwidth]{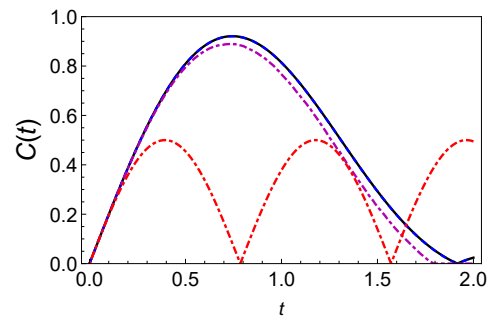}
    \caption{The smooth black curve here shows the concurrence between the first and second spin of an Ising spin chain in the presence of a transverse field for $N=7$ while the time effective XY spin chain is shown by the dashed blue curve. The dot-dashed magenta curve shows the effect of noise while the dotted red curve is for the Ising spin chain in the absence of fields. Here, we have set the neighboring coupling strengths as $J_i=1$.}
    \label{fig4}
\end{figure}

We first consider the concurrence between the first and second spins of the Ising spin chain in the presence of a transverse control field as described by Eq.~(\ref{eq4}). Results are shown in Fig.~(\ref{fig4}). The smooth black curve is for the Ising spin chain in the presence of the transverse field while the dashed blue curve is for the time-effective XY spin chain. The magenta curve shows the effect of noise which we find to match closely, indicating that the effect of the noise is largely mitigated by our control field. The dotted red curve is for the Ising spin chain in the absence of any fields, and we see that this is significantly lower. These results demonstrate the usefulness of using our simple control field to engineer the spin chain Hamiltonian so as to generate greater entanglement.   

\subsection{The XXX spin chain}
Earlier, in Sec.~(\ref{Sec_2_a_2}), we saw that we can arrive at the XXX spin chain by choosing $J_i^x=2$, $J_i^y=1$, and $A=0$ in Eq.~(\ref{eq12}) which we obtained by  applying a control field along the $y$ direction to the XY spin chain. The choice of setting $A=0$ again implies that $\mathcal{J}(\frac{4g}{\omega})$ should be set to zero and that $\frac{4g}{\omega}$ equals the zeroth roots of the Bessel function. Once again, we set $g=\frac{\omega}{4(2.4048)}$. 

\begin{figure}[h!]
    \centering
    \includegraphics[width=0.35\textwidth]{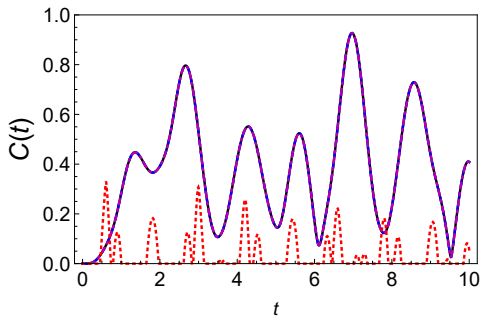}
    \caption{The smooth black curve shows the concurrence of entanglement generation between the ends of the XY spin chain in the presence of the control field along the $y$ axis such that the neighboring coupling strength are set as $J_i^x=2$ and $J_i^y=1$ for $N=5$, the dashed blue curve shows the time effective XXX spin chain, the dot-dashed magenta curve shows the effect of noise on the time-dependent system, while the red dotted curve is for the system in the absence of any fields.}
    \label{fig5}
\end{figure}

\begin{figure}[h!]
    \centering
    \includegraphics[width=0.35\textwidth]{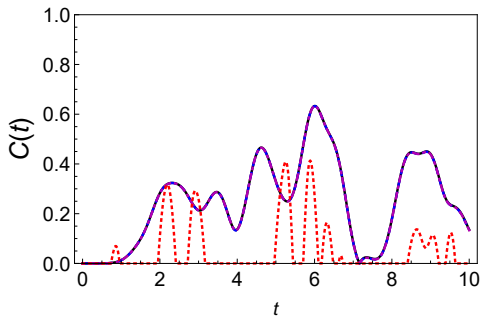}
    \caption{Same as Fig.~(\ref{fig5}) except for $N=8$.}
    \label{fig6}
\end{figure}

We now examine the generation of end to end entanglement in the XXX spin chain for our initially bit flipped state. Figures (\ref{fig5}) and (\ref{fig6}) present the concurrence of entanglement generation between the ends of the XXX spin chain for $N=5$ and $N=8$ respectively. We see that the time-dependent system closely matches with the our effective time-independent treatment since the solid black curve and dashed blue curve coincide. The system is also able to effectively cancel out noise terms as is shown by the dot-dashed magenta curves. At the same time, we see that the concurrence of the initial XY spin chain without any external fields is considerably less - this is demonstrated by the dotted red curve in Figures (\ref{fig5}) and (\ref{fig6}). Once again, these results demonstrate the usefulness of applying our simple control field.

\subsection{The rotated XXZ spin chain}

\begin{figure}[h!]
    \centering
    \includegraphics[width=0.35\textwidth]{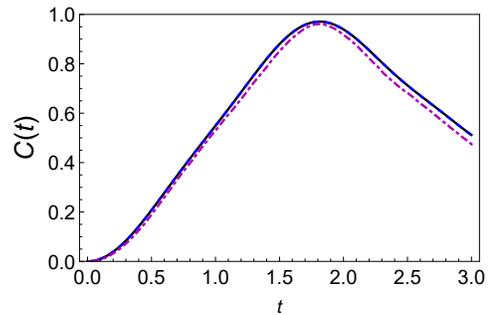}
    \caption{The smooth black shows the concurrence between the first and last spins of an XY spin chain in the presence of the control field along the $y$ axis for $N=3$. The dashed blue curve shows the effective rotated XXZ spin chain while the magenta dot-dashed curve shows the effect of noise. Here, we have taken $J_x^i=J_y^i=1$.}
    \label{fig7}
\end{figure}

\begin{figure}[h!]
    \centering
    \includegraphics[width=0.35\textwidth]{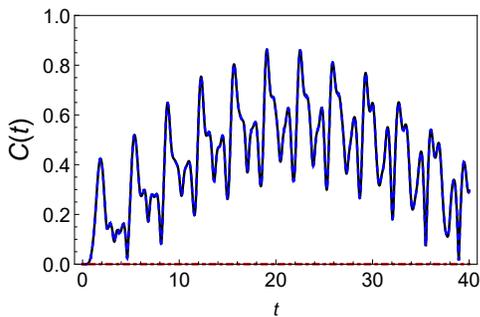}
    \caption{Same as Fig.~(\ref{fig7}) except for $N=8$ and $J_i=2$. The dot-dashed red curve tracing the horizontal axis is for the XY spin chain without any fields.}
    \label{fig8}
\end{figure}

We now present results for the generation of entanglement between two ends of a rotated XXZ spin chain as found earlier in Eq.~(\ref{eq14}). We recall that we applied the control field along the $y$ axis to the XY spin chain and chose our parameters such that $J_x^i=J_y^i=J_i=1$ and set $A=0$ in Eq.~(\ref{eq12}). For this case too, we initialize the state as the bit-flipped state; however, since we have the rotated XXZ spin chain, we now consider the initial state to be $\ket{\Psi_i}=\ket{0}_y\otimes\ket{1}_y\ldots\otimes\ket{1}_y$. Moreover, as the control field is along the $y$ axis this time, it can only decouple noise terms along the $x$ and $z$ directions. Hence, we to show the usefulness of our control field in eliminating noise, we now consider $H_{SB}=\sum_{i=1}^N (B_x^i\sigma_x^i +B_z^i\sigma_z^i)$. We have found significant entanglement generation between the ends of a rotated XXZ spin chain by setting the neighboring interactions equal. This is demonstrated in Figures (\ref{fig7}) and (\ref{fig8}) for $N=3$ and $N=7$ respectively where the smooth black curve is for the XY spin chain in the presence of a control field along the $y$ axis while the dashed blue curve is for the time-effective rotated XXZ spin chain. These practically coincide. The magenta curve demonstrating the effect of noise for the time-dependent system shows that the control field significantly cancels out the noise terms. More so, the dot-dashed red curve in Fig.~(\ref{fig8}) shows the nearly non-existent concurrence in the XY spin chain without the field, which we have found this to be the case for other values of $N$ too. We have investigated and found the same qualitative behavior for increasing values of $N$.

\subsection{Ising spin chain with nearest and next nearest neighbor interactions}

\begin{figure}[h!]
    \centering
    \includegraphics[width=0.35\textwidth]{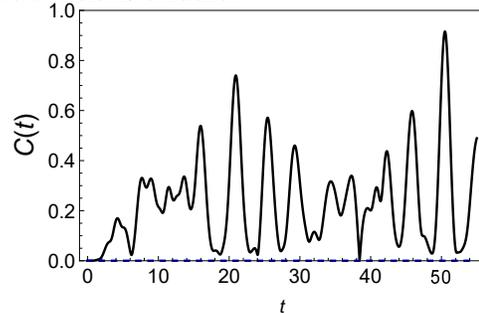}
    \caption{The smooth black curve shows the concurrence for the entanglement generated between the ends the Ising spin chain with next nearest neighbor interactions in the presence of the control field along the $z$ direction for $N=9$ while the dashed blue curve is for the system without any fields. Here, we have set $J_i=L_i=1.$}
    \label{fig9}
\end{figure}

So far, we have modeled spin chains taking into account only the nearest neighbor interactions. A more realistic model however would take into account interactions beyond the nearest neighbor. As such, we take into account the next nearest neighbor interactions and modify the Ising spin chain Hamiltonian in the presence of a transverse sinusoidal field as

\begin{equation}
    H(t) = \sum_{i=1}^{N-1}J_{i}\sigma_x^{i}\sigma_x^{i+1} + \sum_{i=1}^{N-2}\frac{L_i}{N}\sigma_x^{i} \sigma_x^{i+2} \sum_{i=1}^{N}h(t)\sigma_z^{i},
\label{eq21}
\end{equation}

where, as before,  $h(t) = g\cos(\omega t)$. Again, we transform Eq.~(\ref{eq21}) to the interaction picture and find that it yields the time-effective Hamiltonian 
\begin{eqnarray}
    H_{I}= \sum_{i}^{N-1}\frac{J_i}{2}[(A+1)\sigma_{x}^{i}\sigma_{x}^{i+1}-(A-1)\sigma_{y}^{i}\sigma_{y}^{i+1}]+\nonumber\\
    \sum_{i}^{N-2}\frac{L_i}{2N}[(A+1)\sigma_{x}^{i}\sigma_{x}^{i+2}-(A-1)\sigma_{y}^{i}\sigma_{y}^{i+2}]. 
    \label{eq22}
\end{eqnarray}
The result derived in Eq.~(\ref{eq22}) is analogous to the transformation of the Ising spin chain with only nearest neighbor interactions in the presence of a transverse field as shown earlier in Eq.~(\ref{eq6}) - we have transformed the Ising spin chain with nearest and next nearest neighbor couplings to the completely anisotropic XY spin chain with nearest and next nearest neighbor couplings. If we choose $A=0$ in Eq.~(\ref{eq22}), we find that this yields the completely isotropic XY spin chain with both nearest and next nearest neighbor interactions. Let us then present results for the concurrence of the entanglement generated between the ends of the XY spin chain with both nearest and next nearest neighbor interactions. As usual, this time-independent spin chain corresponds to the time dependent system in Eq.~(\ref{eq21}) with $g = \frac{\omega}{4(2.4048)}$ for the time-dependent system. Results for the concurrence of end to end entanglement generation for $N=9$ are presented in Fig.~(\ref{fig9}). Here the smooth black curve shows the concurrence for the time-independent effective spin chain in Eq.~(\ref{eq22}) while the dashed blue curve tracing the $x$ axis is for the Ising spin chain with nearest and next nearest without any fields. Once again, the drastic advantage of our control field is evident. 

\section{Conclusion}

In summary, we have applied a continuous driving field to spin chains to effectively generate additional spin-spin interactions. These interactions can be tuned; in particular, interaction strengths depend on Bessel functions of the first kind, and can be physically changed by varying the strength of the driving field. In particular, we have considered how we can obtain the isotropic XY spin chain from the quantum Ising model, as well as the XXX and XXZ spin chains from the XY model. We have also briefly extended our results to include next-nearest neighbor interactions. Moreover, our control field has been shown to also mitigate the effects of noise on the spin chain. The particular cases we have analyzed yield interesting results from the point of view of quantum information processing tasks. While the isotropic XY spin chain yields perfect state transfer, the other time effective spin chains help us generate significantly high entanglement as compared to the case where no control field is applied.

\providecommand{\noopsort}[1]{}\providecommand{\singleletter}[1]{#1}%

\end{document}